\documentclass[paper]{JHEP}
\usepackage{amsmath}
\usepackage{amssymb}
\usepackage{epsfig}

\def\cO#1{{\cal{O}}\left(#1\right)}

\def\MSbar{\overline{\mbox{\scriptsize MS}}}

\def\half{\mbox{\small $\frac{1}{2}$}}
\def\cR{{\cal{R}}}

\def\cf{C_F}
\def\ca{C_A}
\def\nf{n_f}

\def\la{\mathrel{\mathpalette\fun <}}

\def\fun#1#2{\lower3.6pt\vbox{\baselineskip0pt\lineskip.9pt
  \ialign{$\mathsurround=0pt#1\hfil##\hfil$\crcr#2\crcr\sim\crcr}}}

\def\beq{\begin{equation}}
\def\eeq{\end{equation}}

\def\beql#1{\begin{equation}\label{#1}}
\def\beeq{\begin{eqnarray}}
\def\eeeq{\end{eqnarray}}
\def\bit{\begin{itemize}}
\def\eit{\end{itemize}}

\def\eps{\epsilon}

\def\as{\alpha_{\mbox{\scriptsize s}}}

\title{On the QCD analysis of Jet Broadening}
\author{Yu.L. \ Dokshitzer\thanks{On leave from St. Petersburg Nuclear
    Institute, Gatchina, St. Petersburg 188350, Russia}, A.\ Lucenti,
  G.\ Marchesini and G.P.\ Salam \\
  Dipartimento di Fisica, Universit\`a di Milano \\
  and INFN, Sezione di Milano, Italy \\}

\abstract{The perturbative all-order analysis of the jet-broadening
  $B$-distribution in the small-$B$ region is carried out with
  single-logarithmic accuracy, which requires the control of both the
  sum of the moduli and the modulus of the sum of the transverse
  momenta of soft gluons.  We confirm the master equation for the
  $B$-distribution derived by Catani, Turnock and Webber (CTW). Proper
  treatment of quark recoil is necessary at this accuracy. This effect
  was neglected in the CTW solution. We show that the answer can be
  expressed in terms of the CTW result but evaluated at a properly
  rescaled $B$ value.
  }


\preprint{IFUM-602-FT\\
  hep-ph/9801324 \\
  January 1998}

\begin{document}
\psfull




\section{Introduction}

Interesting characteristic features of a given shape variable can 
be appreciated only in calculations to next-to-leading order.  
One of the most
interesting variables is the jet-broadening $B$-distribution,
introduced in Ref.~\cite{Broad}, in which $2B$ is the sum of the
moduli of the transverse momenta of all emitted particles with respect
to the thrust axis in units of $Q$, the total final state mass.

Consider, in perturbative QCD, the radiation emitted in 
$e^+e^-$ annihilation  consisting of the primary quark and antiquark 
with 4-momenta $p$ and $\bar p$ and the secondary partons $k_i$.
Hereafter we shall consider small values of $B$ so that the secondary 
partons are soft and the primary $p$ and $\bar{p}$ belong to opposite 
hemispheres. 
To be specific we define the right hemisphere as the one containing 
the quark $p$.

The simplest variable is the single-jet broadening 
(the right-jet broadening $B_R$ in what follows) defined by
\begin{equation}
\label{BRdef}
 2B_R  \;=\; \sum_{i\in R}\lvert \vec{k}_{t i}\rvert\;+\;p_t\>,
\qquad \vec{p}_t\;=\; -\sum_{i\in R} \vec{k}_{ti} \,.
\end{equation}
Since the transverse momenta are taken with respect to the thrust axis,
the total vector sum of transverse momenta in each hemisphere is zero.
 One introduces the total jet-broadening $B_T$, as the sum of right- and 
left-jet broadening ($B_T=B_R+B_L$), and the wide-jet broadening 
$B_W=\mbox{Max}\{B_R,B_L\}$.

We consider first the $B_R$-distribution.  The jet-broadening
distribution for the right hemisphere is given in terms of the
multi-parton emission distribution $d\sigma_n$ by
\begin{equation}
  \label{sBR}
\begin{split}
  \frac{d\sigma}{\sigma (d\ln B_R)} =&
  \frac{d}{d\ln B_R}\> I_R(B_R)\>,\\
  I_R(B_R)=& 
  \sum_n \int \frac{d\sigma_n}{\sigma} \>\> \Theta\left( 2B_R
  -\sum_{i\in R} \left|\vec{k}_{ti} \right|
  -\left|\sum_{i\in R} \vec{k}_{ti} \right| \right) \,.
\end{split}
\end{equation} 
For small $B_R$, which corresponds to all the final-state partons 
having small transverse momenta, one can approximate $d\sigma_n$ as a 
product of two factors. 
The first is a coefficient factor which depends only on $\as(Q)$. 
The second is an evolutionary exponent, which describes the 
production of small-$k_t$ partons off the primary quark-antiquark pair.  
Hence
\begin{equation}
  \label{universal}
  I_R(B) = C(\as(Q)) \Sigma_R(B,\as) \,.
\end{equation}
The essential momentum scales in the coupling in $\Sigma_R$ range from
$B_R Q$ to $Q$. The perturbative treatment that we shall pursue
requires $B_R Q \gg \Lambda_{QCD}$. The accuracy of the perturbative
treatment is limited by non-perturbative power corrections of relative
order $(\ln Q) \Lambda_{QCD}/(B Q)$ which are treated elsewhere
\cite{???,DLMSuniv}.

If only one gluon is present then the quark and gluon transverse
momenta are equal and opposite --- hence the natural factor of two in
the definition of $B_R$ \eqref{BRdef}. But starting from two gluon
emission, the situation already becomes significantly more complex,
because the modulus of the quark transverse momentum depends not only
on the moduli of the gluon transverse momenta but also on their
relative angles.  So in higher orders one has to control
simultaneously the sum of the moduli of the momenta, and the modulus
of their vector sum.  It turns out that in the kinematical region
where $\as\ln 1/B_R \ll 1$, one gluon has a transverse momentum much
larger than that of the others, so that the quark contribution is
still $\half p_\perp\simeq \half B_R$ and therefore easily accounted
for~\cite{CTW}. On the other hand, in the region of extremely small
$B$, $\as\ln 1/B_R \gtrsim 1$, the gluonic contribution to $B_R$ comes
from several gluons with comparable transverse momenta, so that the
problem of an accurate treatment of the quark recoil becomes severe.

A perturbative analysis based on the all-order resummation of leading
and next-to-leading logarithmically enhanced contributions to
jet-broadening distributions was
performed by Catani, Turnock and Webber (CTW) in \cite{CTW}. Their
result in the kinematical region $B \ll 1$ is based on the
``exponentiation'' of one gluon emission. The approach that they
developed guarantees that in the corresponding exponent, all terms
with $\as^m \ln^n 1/B, m\le n$ are kept track of. At a given point,
they made the approximation that the quark recoil contribution to
$B_R$ is always $\half p_\perp = \half B_R$. This simplifies the
answer, but mistreats the $\as^n\ln^n 1/B$ terms, starting from $n=2$.
In this note we will show how to treat properly the quark recoil and
thus how to improve the
CTW prediction for the jet-broadening spectrum.

Before discussing our result 
we first recall the form of the double-logarithmic contribution.  
Here one can simplify the analysis by assuming $\half p_t = \half B_R$.
Moreover one assumes that all the final-state partons are soft and
collinear gluons, emitted independently.  This small $B_R$
contribution to the jet-broadening distribution is given by
\begin{equation}
  \label{BRlead}
    \Sigma_R(B_R)= e^{-R_0(1/B_R)}\;(1+\cO{\as^n\ln^n 1/B_R})\,,
\end{equation}
where $R_0(1/B)$ is the soft part of the gluon emission 
distribution (see Appendix~A) integrated over the region $k_t>B_R$
\begin{equation}
  \label{R0}
  R_0(1/B_R)
  \;=\; \int_{Q^2B^2_R}^{Q^2}\frac{dk^2_{t}}{k_{t}^2} 
  \int_{k_t}^1 dz
  \;\frac{\as(k_t) \cf}{2\pi}\; \frac{2}{z}
  \;=\; \frac{\as(Q) \cf}{\pi}\;\ln^21/B_R + \ldots
\end{equation}
The limitation $k_t>B_R$ comes from the fact that for small $B_R$,
to leading order the real emission takes place only for $k_t<B_R$ and
here is cancelled by part of the virtual contribution.  
Thus only the virtual contribution remains in the region $k_t>B_R$. 
Taking into account the running of the coupling, $R_0$ becomes a
series with terms of the form $\as^n \ln^{n+1} B$, while the neglected
terms are one power of $\ln B$ down. 

In our analysis, we intend to compute all single-logarithmic
corrections, i.e.\ corrections of order $\as^n \ln^n B$. 
We show that to achieve this accuracy it suffices to
``exponentiate'' the next-to-leading order single-gluon emission
formula \cite{CTW} and to treat properly the quark recoil. 
The final result can be written in the form:
\begin{equation}
  \label{BRnext}
 \begin{split}
  \Sigma_R(B_R)&\>=\>\left(\frac{e^{-\gamma_ER'}}{\Gamma(1-R')}\right)
  \>e^{-R\,(\lambda/2B_R)}\;
  (1+\cO{\as^n\ln^{n-1}1/B_R})\,,
  \quad n \ge 2\,,\\
   & R'\>\equiv -\frac{\partial R(1/B_R)}{\partial \ln B_R}
 \end{split}
\end{equation}
where the radiator $R(\lambda/2B_R)$ is given by the one-gluon emission
distribution, as in \eqref{R0}, but obtained from the next-to-leading order
splitting function.
The argument of the radiator is the jet-broadening $B_R$ 
rescaled by a function $2/\lambda$ 
which has a single-logarithmic expansion, i.e.\ with leading terms 
of the order $\as^n \ln^n 1/B_R$.  
For $B_R$ finite, $\lambda$ tends to $2$, while for small $B_R$ it 
tends to $1$. 
The function $\lambda$ takes into account the effect of quark recoil. 
If one neglects quark recoil and puts for instance the quark
transverse momentum at $p_t=B_R$, as done in \cite{CTW}, then 
one finds exactly the same expression \eqref{BRnext} but with 
the function $\lambda$ frozen at $\lambda=2$. 

Within our calculation we are also able to show the absence of the
first non-logarithmic correction to $I(B)$ that comes from the running
coupling at low scales, $\as(B Q)$. Though formally classified as
being subleading, $\cO{\as^n\ln^{n-1}1/B}$, numerically it could be
dangerous, since $\as(B Q)$ increases at small $B$.

It may be surprising that the accuracy in \eqref{BRnext} can be
achieved by using an independent multi-gluon emission distribution.
The important point to realise is that subsequent gluon decay can be
neglected to single-logarithmic order. This will be discussed in detail in
the paper. However the following simple argument explains why.  The
$B$-spectrum contains a characteristic exponent of $\as\ln^21/B_R$.  A
value of $B$ can be changed by a non-collinear non-soft gluon decay, 
with the transverse momenta of both offspring partons being $\cO{BQ}$.
Given the relative probability of such a decay, $\cO{\as}$, we get a
correction $B\to B(1+\as)$, which translates into
$\as\ln^2(B(1+\as))=\as\ln^2B+\as^2\ln B$, the latter being a
negligible effect with a power of the $\log$-enhancement factor
smaller than that of $\as$.

The paper is organised as follows. In Sect.~2 we derive the $B_R$
distribution using the independent emission distribution and by
including the effect of quark recoil.  We generalise the analysis to
the total and wide jet-broadening distributions $\Sigma_T(B)$ and
$\Sigma_W(B)$.

In Sect.~3 we show that our result can be derived from the CTW
equation obtained from the coherent branching process.  To achieve the
desired single-logarithmic accuracy one has to improve the treatment
of recoil. We also verify that one can neglect the gluon branching
process, thus justifying the use of the independent gluon emission
approximation.

In Sect.~4 we perform the actual evaluation of $\Sigma(B_R)$ to the
required accuracy.

In Sect.~5 we present the results of numerical analysis. We compare
the final result with $\cO{\as^2}$  numerical results obtained 
with the EVENT2 program \cite{CS}. We give all the necessary
information to be able to carry out the matching with the fixed order
perturbative results. We show the difference between our result and
that of CTW, and compare the matched and non-matched calculations.

In Sect.~6 we discuss, comment and look forward.

\section{Independent gluon emission}
To compute jet broadening to single-logarithmic accuracy, it is enough
to use the independent gluon emission distribution from the primary
quark-antiquark pair. This is given in terms of the $q\to q+g$
splitting function to next-to-leading accuracy.

The independent gluon emission distribution is 
\begin{equation}
  \label{wn}
  dw_n  \;=\;
  \frac1{n!} \prod_{i=1}^n \frac{d^2k_{t i}}{\pi k_{t i}^2}
  dz_i\;2P_{qq}[\as,z_i]\; \Theta(z_i - k_{t i}) \, V\,,
\end{equation}
where the factor 2 takes into account the fact that, for each gluon,
the integration domain includes both the right and left hemispheres.
The quark splitting function is given, in the $\MSbar$ scheme by
\begin{equation}
\label{PqqMS}
\begin{split}
  P_{qq}[\alpha_{\MSbar},z] &=
 \cf\frac{\alpha_{\MSbar}(k_t)}{2\pi}\;\frac{1+z^2}{1-z}
 \left[1\;+\;\frac{\alpha_{\MSbar}}{2\pi}\;K \right] + \cdots
 \\
 K & = \ca\left(\frac{67}{18}-\frac{\pi^2}{6}\right)-\frac{5}{9}n_f\,,
\end{split}
\end{equation}
for $n_f$ flavours. The non-soft part of the quark splitting function,
$1-z\sim1$, produces a single-logarithmic contribution to the exponent,
and must be kept, while the neglected part of the two-loop anomalous
dimension generates negligible terms of order $\as^2\ln B$.
The phase space region is $(1-z)>k_t/Q$ (see Appendix~A).

We work in the physical (CMW) scheme~\cite{CMW} in which $\as$ is the
measure of the intensity of soft emission. We then have
\begin{equation}
    \label{Pqq}
  P_{qq}[\as,z] \,=\,\cf \frac{\as(k_t)}{2\pi}\;\frac{1+z^2}{1-z}\,,
  \quad 
  \as(k_t) = \alpha_{\MSbar}(k_t)
  \left(1\;+\;\frac{\alpha_{\MSbar}}{2\pi}\;K \;+\;
  \cO{\alpha^2_{\MSbar}}\right)\,.
\end{equation}
To this order, the virtual correction $V$ is given by
\begin{equation}\label{virt}
  \ln V \,=\,
  -\int^{Q^2} \frac{dk_t^2}{k_t^2}\int_0^{1-k_t/Q}
  dz\,2P_{qq}[\as,z] \,. 
\end{equation}
For collinear-safe inclusive quantities one can integrate over
$k_t$ in \eqref{virt} and $k_{t i}$ in \eqref{wn} down to zero.

The distribution \eqref{wn} is normalised to unity: 
$\sum_n \int dw_n =1$, where the transverse momentum of each real gluon is
integrated up to $Q$.
The independent emission distribution is valid only for small $k_{ti}$.
It simplifies the treatment but mistreats the non-logarithmic region
of large transverse momenta, $k_t\sim Q$, both in real and virtual
terms. 
This is compensated by the factor $C(\as)$ in \eqref{universal}.
For small $B$ values this factor is $B$-independent. The
$B$-dependence is embodied into the $\Sigma$ factor,
\begin{equation}
  \label{SRind}
  \Sigma_R(B) =  \sum_n \int \> dw_n \>\> \Theta\left( 2B
  -\sum_{i\in R} \left|\vec{k}_{ti} \right|
  -\left|\sum_{i\in R} \vec{k}_{ti} \right| \right) \,.
\end{equation}
To make use of the factorisation structure of the multi-gluon matrix
element \eqref{wn} 
we introduce the Mellin representation of the $\Theta$-function 
\begin{equation}
\label{MellFour}
  \Theta\left( 2B
  -\sum_{i\in R} \left|\vec{k}_{t i}\right|
  -\left|\sum_{i\in R} \vec{k}_{t i} \right| \right)
  \;=\; \int \frac{d\nu}{ 2\pi i\nu} \>e^{2\nu B}\>
  \frac{d^2p_{t}\; d^2 b}{(2\pi)^2}  \> e^{-i \vec{p_t}\cdot\vec{b}}
  \> e^{-\nu p_t} \prod_{i\in R}
  e^{-\nu k_{t i}} e^{-i\vec{b}\cdot\vec{k_{t i}}}\,.
\end{equation}
Here $\nu$ is the Mellin variable conjugate to $B$, which runs
parallel to the imaginary axis, to the right of $\nu=0$. 
We have introduced the integration over the quark transverse 
momentum $\vec{p}_t$ and the constraint 
$\vec{p}_t=-\sum_{i\in R} \vec{k}_{t i}$ is implemented by the integration
over $\vec b$. 
We obtain
\begin{equation}
\label{SigmaR}
  \Sigma_R(B)\;=\;
  \int \frac{d\nu}{2\pi i \nu} e^{2\nu B} \;\sigma(\nu)\,,
\end{equation}
where the contour lies to the right of all singularities
of $\sigma(\nu)$ which is given by
\begin{equation}
\label{sigma}
  \sigma(\nu)\;=\;
  \int \frac{d^2p_t d^2 b}{(2\pi)^2} e^{-i \vec{p_t}\cdot\vec{b}}
  e^{-\nu p_t} \cdot e^{-\cR\,(\nu,b)} \;=\;
  \int_0^\infty \frac{\nu bdb}{(\nu^2 + b^2)^{3/2}} \; e^{-\cR(\nu,b)}\,,
\end{equation}
with the radiator
\begin{equation}
\label{cR}
    \cR(\nu,b)\;=\;
    \int\frac{d^2k_t}{\pi k_t^2}
     dz \;P[\as,z]\;
    \left[\,1\;-\;e^{-\nu k_t} e^{-i\vec{b}\cdot\vec{k_t}}\,\right]\,.
\end{equation}
In the CMW scheme the two-loop $\as^2$ contribution to the quark splitting
function \eqref{Pqq} is regular at $z=1$ and can be neglected because
it produces corrections to \eqref{SRind} of order
$\as^n \ln^{n-1}1/B$.
In the $\MSbar$ regularisation scheme instead one should explicitly keep 
the infrared-singular contribution of order $\alpha^2_{\MSbar}$ 
to the splitting function \eqref{PqqMS}, in order to reach 
the desired accuracy, as done in~\cite{CTW}.

Notice that one is completely inclusive with respect to gluons 
emitted in the left hemisphere. 
Thus their contribution cancels $\sqrt{V}$, i.e.\
half of the virtual corrections.
The remaining $\sqrt{V}$ corresponds to the term $1$ in the square bracket
of \eqref{cR}.

Performing the azimuthal integration we obtain
\begin{equation}
\label{cR'}
 \cR(\nu,b) =  \int_0^1\frac{dk_t}{k_t}\>\phi(1/k_t)
  \left[1-e^{-\nu k_t} J_0(b k_t)  \right]\,,
\end{equation}
with $\phi(1/k_t)$ the one-gluon radiation formula to next-to-leading
accuracy
\begin{equation}
\label{phi}
  \phi(1/k_t) \,=\,2\frac{\as(k_t)\cf}{\pi}\;
  \left(\ln \frac{1}{k_t}-\frac34\right)\>+\> \cO{\as^2(k_t)}\,.
\end{equation}
In the leading double-logarithmic approximation $\Sigma_R(B)$
is obtained as follows. 
Since quark recoil is irrelevant to this order, we are free to set 
$p_t=B_R$, which corresponds to neglecting the $b$-dependence 
in the radiator, i.e.\ replacing $\cR(\nu,b)$ with  
$\cR(\nu,0)$ in \eqref{sigma}.
Then, for large $\nu$, one approximates
$[1 - e^{-\nu k_t}] \to \Theta(k_t-1/\nu)$ and further approximates 
the $\nu$-integration by setting $\nu \to 1/B$. 
This leads to the leading order result \eqref{BRlead}.

Following CTW, from $\sigma(\nu)$ one deduces 
the total jet broadening $\Sigma_T(B)$ and 
the wide-jet broadening $\Sigma_W(B)$
distributions given, for small $B$, by
\begin{equation}
\label{SigmaTW}
  \Sigma_T(B)\;=\; \int \frac{d\nu}{2\pi i \nu} e^{2B\nu} \;\sigma^2(\nu)
  \,,
  \quad
  \Sigma_W(B)\;=\;
 \left(\int \frac{d\nu}{2\pi i \nu} e^{2B\nu} \;\sigma(\nu)\right)^2
  \,.
\end{equation}
Characteristic values of the Laplace parameter $\nu$ and of the impact
parameter $b$ that determine the final answer for the $B$-distribution
satisfy
$ \nu\cdot BQ \sim 1\>, \quad b\sim\nu$, 
hence, in the $B \ll 1$ kinematical region we have a large parameter
$   \nu Q\sim 1/B \>\gg\>1$.

A systematic evaluation of both $\sigma(\nu)$ and $\Sigma(B)$ to the
required accuracy will be considered later. 
First we discuss the connection to the approach of CTW, 
obtained in the framework of coherent branching.

\section{Coherent branching}
In \cite{CTW} a technique for analysing $\Sigma(B)$ was
developed, based on the evolution equation for  
the distribution $T_q(Q,\vec{k},P_t)$ of the variable
$$
P_t= \sum_{i\in R} \left|\vec{k}_{ti} \right|
    +\left|\sum_{i\in R} \vec{k}_{ti} \right|
$$
in a quark jet produced with vector transverse momentum $\vec k$.
The final physical distribution is obtained by setting $\vec k=0$. 
The distribution  $\sigma(\nu)$ in \eqref{sigma} is given by 
its Laplace transform at $\vec k=0$
\begin{equation}
  \label{CTWlaplace}
  \sigma(\nu)\;=\;\tilde T_q(Q,\vec 0,\nu)\,,
  \qquad
  \tilde T_q(Q,\vec k,\nu)\;=\; \int_k^\infty
  dP_t\;e^{\nu(k-P_t)}\;T_q(Q,\vec k,P_t)\,.
\end{equation}
 From the coherent branching picture one has the following evolution
equation (see \cite{CTW})
\begin{multline}
\label{CTWLapl}
  \tilde{T}_q(Q,\vec{k};\nu) = 1 + \int_0^Q\frac{d^2\tilde{q}}
   {\pi\,\tilde{q}^2}\int_0^1dz\,P_{qq}[\alpha_s,z]
   \left\{e^{\nu(|\vec{k}|-|z\vec{k}+\vec{q}_t|
-|(1-z)\vec{k}-\vec{q}_t|)} \right. \cdot \\
\left.
\tilde{T}_q(z\tilde{q},z\vec{k}+\vec{q}_t;\nu)
\tilde{T}_g((1-z)\tilde{q},(1-z)\vec{k}-\vec{q}_t;\nu)
-\tilde{T}_q(\tilde{q},\vec{k};\nu)\right\},
\end{multline}
where 
$\vec{q}_t \equiv z(1-z)\vec{\tilde{q}}$, $q_t=|\vec{q}_t|$
and $k=|\vec{k}|$. Here, $\tilde{T}_g$ is the corresponding
distribution for a gluon jet.
The first term in the curly brackets describes real parton
splitting, while the second subtraction term accounts for the virtual effects.

Introducing the Fourier transform
\begin{equation}
  \Gamma_i(Q,\vec b, \nu)= 
  \int \frac{d^2k}{2\pi} \;e^{i\vec b \vec k \;-\; \nu k}\>
  T_i(Q,\vec{k},\nu)\,,
  \quad i=q,g\,,
\end{equation}
we find 
\begin{multline}
  \label{Gammaeqbgen}
  \Gamma_q(Q,\vec{b},\nu)
   = N(\nu,b)+  \int_0^{Q^2}\frac{d\tilde{q}^2}
   {\tilde{q}^2}\int_0^1dz\,P_{qq}[\alpha_s,z]\\
   \left\{ \int \frac{d^2r}{2\pi} J_0(rz(1-z)\tilde{q})\,
   \Gamma_q(z\tilde{q},\vec{b}+(1-z)\vec r,\nu)\,
   \Gamma_g((1\!-\!z)\tilde{q},\vec{b} -z\vec{r},\nu)
 \> -\>\Gamma_q(\tilde{q},\vec{b},\nu)\right\},
\end{multline}
where $N(\nu,b)$ is the Fourier transform of the inhomogeneous term
\begin{equation}
    \label{Nterm}
  N(\nu,b) =  \int \frac{d^2k}{2\pi}\> e^{i\vec{b}\vec{k}-\nu k}
           =  \frac{\nu}{(\nu^2+b^2)^{3/2}}\,.
\end{equation}
Since the dependence on the evolution parameter $Q$ is contained in
the upper limit of the $\tilde{q}$--integration, we consider the
logarithmic derivative of \eqref{Gammaeqbgen}:
\begin{equation}
  \label{Gderiv}
  \Gamma'_q(Q,\vec{b},\nu) \>\equiv\> \frac{\partial}{\partial \ln Q} 
  \Gamma_q(Q,\vec{b},\nu)\,.
\end{equation}
Then $\tilde{q}$ in the integrand gets replaced by $Q$
and $ q_t = z(1-z)Q$.
It is straightforward to verify that  
the integral term possesses the damping factors
$
 e^{-\nu\, z(1-z)Q}\>J_0({b}\,z(1-z){Q})\,,
$
and therefore is concentrated at small values of $(1-z)$,
\begin{equation}
\label{concentr}
(1-z)\la \min\left\{ (\nu Q)^{-1}, (bQ)^{-1}\right\}.
\end{equation}
As a result, any correction 
proportional to $(1-z)$, or $\ln z$, will produce a {\em power-suppressed}\/
contribution $\sim 1/\nu Q\propto B\ll 1$.
Taking into account that $r\sim b\sim\nu$, this allows us to 
approximate
$
 q_t = z(1-z)Q \>\simeq\> (1-z)Q, 
$
which correction is of the order of $\as(Q)/\pi$ and will be absorbed
into the coefficient function $C(\as)$ in \eqref{universal}.
We can also replace in \eqref{Gammaeqbgen}  
$$
 \Gamma_q(zQ,\vec{b}+(1-z)\vec{r},\nu) \>\Rightarrow\>  
 \Gamma_q( Q,\vec{b}             ,\nu)\,,
$$
and
$$
 \Gamma_g((1\!-\!z)Q,\vec{b}-z\vec{r},\nu) \>\Rightarrow\> 
 \Gamma_g((1\!-\!z)Q,\vec{b}- \vec{r},\nu)\,.
$$
The accuracy of these approximations is of the order of the 
relative corrections ${\as}/{\pi}\cdot B$.
Notice that the first argument in $\Gamma_g$, generally speaking,  
cannot be expanded in $(1-z)\ll1$ since the corresponding dependence 
is double-logarithmic.

By making these simplifications ($q_t=(1-z)Q$)  
we obtain 
\begin{equation}
  \label{Gammaprimesimp}
 \frac{ \Gamma'_q(Q,\vec{b},\nu)}{\Gamma_q(Q,\vec{b},\nu)}
=2\int_0^1dz\, P_{qq}[\alpha_s,z]
 \left\{ J_0(bq_t)\, e^{-\nu q_t}\,T_g(q_t,q_t,\nu)
  -1\right\}.
\end{equation}
The final solution reads 
\begin{equation}
\begin{split}
  \label{bsol}
\tilde{T}_q(Q,0,\nu)
 &= \int \frac{d^2b}{2\pi}\> N(\nu,b)\, e^{-S(\nu,b)} \>, \\
S(\nu,b) &=
 \int_0^{Q^2}\frac{d^2 q_t}
   {\pi q_t^2}\int_0^{1-q_t/Q}dz\,P_{qq}[\alpha_s(q_t),z]
\left[\,1\;-\;e^{-\nu q_t+i\vec{b}\vec{q}_t}\, T_g(q_t,q_t,\nu)\,\right].
\end{split}
\end{equation}
Neglecting gluon branching, $T_g\Longrightarrow 1$, we obtain the
result of previous section 
\begin{equation}
  \label{CTWmod}
  \sigma(\nu)=\tilde{T}_q(Q,0;\nu)= \eqref{sigma}\,.
\end{equation}
To see that gluon branching, $T_g\neq 1$,
is indeed negligible within our accuracy, 
we look at the correction to the exponent in \eqref{bsol}:
\begin{equation}
  \label{Scorr}
  \delta S(\nu,b) =  \int_0^{Q^2}\frac{d^2 q_t}
   {\pi q_t^2}\int_0^{1-q_t/Q}dz\,P_{qq}[\alpha_s(q_t),z]
\>e^{-\nu q_t+i\vec{b}\vec{q}_t}\, \left[\,T_g(q_t,q_t,\nu) -1 \,\right].
\end{equation}
First we notice that the exponential factor forces $q_t\la \nu^{-1}$. 
On the other hand, 
by examining the evolution equation \eqref{CTWLapl} we observe that 
$T_g(q_t,q_t,\nu)-1$ vanishes for $q_t\ll \nu^{-1}$. 
As a result, the $q_t$-integral is concentrated at $q_t\nu \sim1$,
$
   T_g(q_t,q_t,\nu)-1\> \sim \> \as\> \delta(\ln(q_t\nu))\>,
$
and the correction amounts to 
$$
\delta S \>\sim\> \as^2 \ln(\nu Q)\>, 
$$
with the single logarithmic factor emerging from the $z$-integration.

This first perturbative correction is due to
quark $\to$ (quark + two gluons/$q\bar{q}$) pair splitting
processes in which two secondary partons have similar emission angles and
energies of the same order.
It is easy to check that a correction of the same order originates
from non-collinear two-parton production at {\em large}\/ angles
$\Theta\sim1$, and, in particular, of the configuration of partons
falling into opposite hemispheres.  In the present treatment the left
and right jets contribute independently to the event broadening and so
such subleading contributions are not included.

\section{Evaluation of $\Sigma(B)$}

In this section we first evaluate the radiator $\cR(\nu,b)$ and then
perform the $b$-integral in \eqref{sigma} to evaluate
$\sigma(\nu)$. 
For the sake of simplicity in this section we put $Q=1$.

\subsection{Radiator}
To evaluate the radiator for large $\nu$ and $b$ values
we introduce the function
\begin{equation}
\label{R}
R(\mu) \>=\> \int_{\mu^{-1}}^1\frac{dk}{k}\phi(k^{-1}) \>,
\quad \mu=\half(\nu+\sqrt{\nu^2+b^2})\,,
\end{equation}
which corresponds to \eqref{R0} with the next-to-leading
splitting function. With account of the running coupling, it contains 
all terms of order $\as^n\ln^{n+1}\mu$ as well some of the essential
$\as^n\ln^n\mu$ terms.
We write
$$
\cR(\nu,b)= R(\mu) \>+\> \delta R(\nu,b)\,,
$$
and study the correction 
\begin{equation}
  \label{deltar}
    \delta R(\nu,b) \>=
        \> \int_0^{\mu^{-1}}\frac{dk}{k}\phi(k^{-1})
        \left(1-e^{-k\nu}J_0(bk)\right)
        - \int_{\mu^{-1}}^1\frac{dk}{k}\phi(k^{-1})\>e^{-k\nu}J_0(bk)\>.
\end{equation}
It is determined by the integration region $k\sim\mu^{-1}$
and can be written as ($\eps\to0$)
\begin{multline}
  \label{trick}
    \delta R(\nu,b) \>=\>
   \int_0^{1}\frac{dz}{z}z^\eps\> \phi(\mu/ z )
 - \int_0^{\infty}\frac{dz}{z}z^\eps\>e^{-z\nu/\mu}J_0(b z/\mu)
  \>\phi(\mu /z) \\
=    \int_0^{1}\frac{dz}{z}z^\eps\> \phi(\mu/ z)
 - \int_0^{\infty}\frac{dz}{z}z^\eps\>e^{-z\nu/\mu}
  \>\phi(\mu/ z)
-  \int_0^{\infty}\frac{dz}{z}\>e^{-z\nu/\mu}\left(J_0(b z/ \mu)-1\right)
  \phi(\mu/ z)
\>,
\end{multline}
where we have neglected contributions of order of $\exp(-\mu)$.

To evaluate $\delta R$ we expand $\phi(\mu/z)$ in powers of $\ln z$
$$
\phi(\mu/z)= R'(\mu)-R''(\mu)\,\ln z + \ldots
$$
with
\begin{equation}
\begin{split}
  R' (\mu) &\equiv\> \frac{d}{d\ln\mu} R(\mu)\> = \> \phi(\mu)\,,\\
  R''(\mu)& =\> \frac{d}{d\ln\mu} \phi(\mu)\> =\> 
  2\cf \frac{\as(1/\mu)}{\pi} + \cO{\as^2 \ln \mu} \,,\\
  R'''(\mu) &=\> \cO{\as^2 \ln \mu}\,.
\end{split}
\end{equation}
We have that $R'''(\mu)$ is beyond our accuracy since it would lead to
corrections to $\Sigma(B)$ of order $\as^n\ln^{n-1}1/B$.  In
$R''(\mu)$ we keep the correction of the order of the coupling at the
reduced scale $\as(1/\mu)$ and neglect its higher powers in $\as$.  To
our accuracy we could neglect the scale $1/\mu$ in $\as$.  However we
shall keep track of the non-logarithmic corrections proportional to
$\as(1/\mu)$ that emerge in the course of approximate evaluation of the
$b$- and $\nu$-integrals. This enables us to guarantee that no
first-order correction $\as(BQ)$ is present in the final answer.

The contribution to $\delta R(\nu,b)$ proportional
to $R'(\mu)$, the leading order correction, is given by
\begin{equation}
 \label{deltaRprime}
 \delta R^{(1)} = R'(\mu)\cdot\gamma_E \>.
\end{equation}
Having chosen a different definition of $\mu$ would have resulted in
an additional logarithmic contribution.

The non-logarithmic correction, of order $R''(\mu)\sim \as(1/\mu)$, is
\begin{equation}
  \delta R^{(2)}=R''(\mu)\cdot \Delta(\nu,b)\>,
\end{equation}
where (see Appendix~B)
\begin{equation}
\begin{split}
  \Delta(\nu,b)&=  \left\{ \half\left(\ln\frac\mu\nu+\gamma_E\right)^2
  +\half \psi'(1)  +\> 
  \ln\mu\ln\frac{\nu}{\mu}\>+\>c(\nu,b) \right\},
  \\
  c(\nu,b)=& \int_0^{\infty}\frac{dx}{x}\>\ln x\>
  e^{-\nu x}\left(J_0(bx)-1\right)\>,
\end{split}
\end{equation}
We have then 
\begin{equation}
  \cR(\nu,b) = R(\mu) + \delta R(\nu,b)
  = R(\mu e^{\gamma_E}) +  
  R''(\mu)\cdot(\Delta(\nu,b)-\half\gamma_E^2)\, + \cO{R'''}\,.
\end{equation}
Expanding in $\ln\mu/\nu$ we obtain
\begin{equation}
\label{Rapprox}
  \cR(\nu,b) = R(\bar\nu) + R'(\bar\nu)\>\ln\frac{\mu}{\nu} 
  + R''(\bar\nu)\cdot \bar\Delta\left(\frac{\mu}{\nu}\right)\, 
  + \cO{R'''}\,.
\end{equation}
where
\begin{equation}
\bar\nu\equiv \nu\,e^{\gamma_E} \>,   
\end{equation}
and
\begin{equation}
\bar\Delta(\mu/\nu)= 
  \half \psi'(1)+\gamma_E\>\ln\frac{\mu}{\nu}-\ln\nu\ln\frac{\mu}{\nu}
  +c(\nu,b)\,,
\end{equation}
is a function of the ratio $\mu/\nu$ and has no $\ln \nu$ terms
(see Appendix~B).

\subsection{$\sigma(\nu)$}

Now we substitute \eqref{Rapprox} into \eqref{sigma}
and performing the $b$-integral to evaluate $\sigma(\nu)$.
Changing the integration variable to $y=\sqrt{\nu^2+b^2}$
we obtain ($\mu/\nu=\half(1+y)$) 
\begin{equation}
 \label{sigmanu}
  \sigma(\nu)
  =e^{-R(\bar\nu)} \int_1^\infty \frac{dy}{y^2}
  \left(\frac{1+y}{2}\right)^{-R'(\bar\nu)} 
  \left(1-R''(\bar\nu)\>\bar\Delta\left(\frac{1+y}{2}\right)
  +\cO{R'''}\right)\,.    
\end{equation}
To calculate the main contribution we introduce 
the rescaling function $\lambda(R')$ 
\begin{equation}
\label{lambdadef}
  (\lambda)^{-R'} =
  \int_1^\infty \frac{dy}{y^2}
  \left(\frac{1+y}{2}\right)^{-R'}\,.
\end{equation}
To estimate the correction $R''\sim \as$ to first order 
we can neglect the exponent $R'$ since 
$\bar\Delta$ is of order one (it has no $\ln \nu$ contribution). 
It equals (see Appendix~B)
\begin{equation}
  \label{Deltabar}
\int_1^\infty\frac{dy}{y^2}\>\bar\Delta\left(\frac{1+y}{2}\right)
=\half (\psi'(1)+\ln^22)\,.  
\end{equation}
Finally we arrive at 
\begin{equation}
\label{sigma'}
\begin{split}
  \sigma(\nu)
  & =e^{-R(\bar\nu)}\>\left(\lambda\right)^{-R'(\bar\nu)} 
  \>\left[1-R''(\bar\nu)\>\half (\psi'(1)+\ln^22)  
  +\cO{R'''}\right]\\
  & =e^{-R(\bar\nu\lambda)}\>
  \left[1-\half R''(\bar\nu)\>\psi'(1)
  -\half R''(\bar\nu)(\ln^22-\ln^2\lambda)\>  
  +\cO{R'''}\right]\,,
\end{split}
\end{equation}
Since $\lambda$ is a function of $R'(\bar\nu)$,
the leading terms of its expansion are $\as^n \ln^n \bar \nu$.
For small values of $R'(\bar\nu)$, the function
$\lambda(R')$ is close to $2$,
\begin{equation}
\label{closetotwo}
  \lambda(R')
  = 2  - R'(\bar\nu)\left(\frac{\pi^2}{6} - 2\ln^2 2\right)
  + \cO{{R'}^2},
\end{equation}
while for large values of $R'(\bar\nu)$ it approaches unity:
\begin{equation}
 \label{closetoone}
  \lambda(R')
  = 1 + \frac{\ln (R'(\bar\nu)/2)}{R'(\bar\nu)} + \cO{{R'}^{-2}}.
\end{equation}
Perturbatively, $\lambda$ is close to 2, see \eqref{closetotwo}.
Therefore the quantity $R''(\ln^2\lambda-\ln^22)$ in \eqref{sigma'}
is of order $R'''$ and thus negligible.   
We can write
\begin{equation}
 \label{leftwith}
 \sigma(\nu)= e^{-R(\bar\nu\lambda)}
 \left[1-\half \psi'(1)\, R''(\bar\nu) \>+\> \cO{R'''} \right].
\end{equation}
Notice that this estimate is uniform in $R'$. 
Indeed, for large values of $R'$ the $\ln^2\lambda$ term in
\eqref{sigma'} vanishes, see \eqref{closetoone}.
At the same time, the $\ln^22$ term disappears as well, since 
for large $R'$ the integral \eqref{sigmanu} for the correction is
concentrated near $y=1$ where $\bar{\Delta}(1)=\half\psi'(1)$,
instead of \eqref{Deltabar}.

The next-to-next-leading correction
$R''(\bar\nu)\sim\as(Q/\nu)\sim\as(BQ)$ that we kept in
\eqref{leftwith} will cancel, in the first order, against a similar
correction coming from evaluation of the $\nu$-integral.

\subsection{The $\nu$-integral}
Here we finally compute the integral over $\nu$ in \eqref{SigmaR}
to obtain $\Sigma_R(B)$.
To this end we expand $\sigma(\nu)$ around some point $\nu_0$ in powers
of $\ln(\nu/\nu_0)\equiv\ln t$.
Introducing $\eta_0\equiv \bar\nu_0\lambda(R')$ we have,
to the required accuracy
\begin{equation}
\label{expansio}
  \Sigma_R(B)
  =  e^{-R(\eta_0)}  \int \frac{dt}{2\pi i\,t} \>
  e^{2\nu_0Bt}\>t^{-R'(\eta_0)}
  \left[1-\half R''(\bar\nu)\left(\ln^2t+\psi'(1)\right)\;
  +\;\cO{R'''}\right] \,.
\end{equation}
Notice that the corrections coming from differentiation of 
$\lambda(R')$ in the argument of $R(\bar\nu\lambda)$ 
do not contribute:
$\lambda\simeq(\as\ln \bar\nu)^n$, therefore 
$R'\cdot\lambda'\simeq \as^m\ln^{m-1}\bar\nu$, 
escaping our resolution.

Now we have to choose 
the value of $\nu_0$ such as to keep 
the characteristic value of $\ln t$ not large.
The leading term is given by the following basic  integral
\begin{equation}
\int_C \frac{dt}{2\pi i\,t}\> e^{2B\nu_0\,t}\>
t^{-\gamma} = \frac{(2 B\nu_0)^\gamma}{\Gamma(1+\gamma)}\>,
\quad \gamma=R'(\eta_0)\>,
\end{equation}
and we obtain
\begin{equation}
\Sigma_R(B)=e^{-R(\eta_0)}
\left[1-\half R''(\eta_0)\left(\frac{d^2}{d\gamma^2}+\psi'(1)\right) 
\,+\,\cO{R'''} \right]
\frac{(2 B\nu_0)^\gamma}{\Gamma(1+\gamma)}\>,
\end{equation}
The correction proportional to $R''(\eta_0)$ becomes
\begin{equation}\label{bla}
  \left(\frac{d^2}{d\gamma^2}+\psi'(1)\right) 
\frac{(2 B\nu_0)^\gamma}{\Gamma(1+\gamma)}
= \left(\left[\, \ln(2 B\nu_0) - \psi(1+\gamma)\,\right]^2
       +\psi'(1) - \psi'(1+\gamma)\right) 
\frac{(2 B\nu_0)^\gamma}{\Gamma(1+\gamma)}\>.
\end{equation}

\paragraph{Choice of $\nu_0$.}
To ensure the smallness of the correction one has to choose 
$\nu_0$ in such a way that $\ln(2 B\nu_0)$ remains finite. 
By choosing 
\begin{equation}
\label{nuzerodef}
 \nu_0= \frac{1}{2B}e^{\psi(1)}\>, \qquad 
 \bar{\nu}_0\equiv \nu_0e^{\gamma_E} = \frac{1}{2B}\,,
\end{equation}
the first correction of order $R''$ in \eqref{expansio},  
given by \eqref{bla}, becomes uniformly small:
\begin{equation}
  \label{firstcorr}
  -\half R''\left[\,\left( \psi(1) - \psi(1+\gamma)\right)^2
       +\psi'(1) - \psi'(1+\gamma)\,\right]\,.
\end{equation}
It vanishes in the first order, that is for $\gamma=0$.

The parameter $\eta_0$ reads
$$
 \eta_0 \equiv \bar{\nu}_0\, \lambda\>=\> 
  \frac{\lambda}{2}\, B^{-1}\,,
$$
and we arrive at
\begin{equation}
  \label{CTWshift}
 \Sigma_R(B) = e^{-R(\lambda/2B)} \,\frac{e^{-\gamma_E R'}}
{\Gamma(1+R')} \equiv  \Sigma^{(0)}_R(2B/\lambda) \>.
\end{equation}
This first form of the answer 
is the CTW distribution with the important
modification that the argument of the double-logarithmic radiator is 
$2B/\lambda$ instead of simply $B$. 
Since  $R'$ enters only in the prefactor, we can take  
$R'\>=\> R'(1/B)$. 

We can further simplify the answer by expanding $R(\lambda/2B)$  in
powers of $\ln \lambda/2$.   
Neglecting the contribution $\cO{R''}$, we arrive at the second form
of the answer, 
\begin{equation}
\label{CTWfact}
\Sigma_R\>=\> 
\left(\frac2\lambda\right)^{R'}\cdot
 \frac{e^{-\gamma_ER'}}{\Gamma(1+R')}\>
 e^{-R(1/B)} 
 =\left(\frac2\lambda\right)^{R'}\cdot \Sigma^{(0)}_R(B)\>,
\end{equation}
where $R'\equiv R'(1/B)$.
This is the CTW answer modified by the single-logarithmic factor 
$(2/\lambda)^{R'}$.
The two forms of our answer, \eqref{CTWshift} and \eqref{CTWfact} are
both correct (and equivalent) to single-logarithmic accuracy.

Finally our results for the total and wide jet broadenings (given
in a form analogous to \eqref{CTWfact}) are 
\begin{equation}
\label{Tfact}
\Sigma_T\>=\> 
\left(\frac2\lambda\right)^{2R'}\cdot
 \frac{e^{-2\gamma_ER'}}{\Gamma(1+2R')}\>
 e^{-2R(1/B)} 
 =\left(\frac2\lambda\right)^{2R'}\cdot \Sigma^{(0)}_T(B)\>,
\end{equation}
and
\begin{equation}
\label{Wfact}
\Sigma_W\>=\> 
\left(\frac2\lambda\right)^{2R'}\cdot
 \frac{e^{-2\gamma_ER'}}{\Gamma^2(1+R')}\>
 e^{-2R(1/B)} 
 =\left(\frac2\lambda\right)^{2R'}\cdot \Sigma^{(0)}_W(B)\>,
\end{equation}
respectively. As before, $R'\equiv R'(1/B)$. The value of $C(\as)$ in
\eqref{universal} is 
\begin{equation}
  \label{Cval}
C(\as) = 1 + \sum_{n=1} C_n \as^n\,, \qquad
C_1 = \frac{\cf}{2\pi}\left(\pi^2 - \frac{17}{2}\right)\,.
\end{equation}
For consistency with the order at which we have performed the
resummation, it is sufficient to know $C$ only to first order in
$\as$.

\section{Numerical analysis}

\subsection{Comparison with two-loop result}

We compare the single-logarithmic result for the total and wide 
jet distribution ($i=T,W$)
$$
\frac{d\sigma_i}{\sigma (d\ln B)}=\frac{d\> I_i(B)}{d\ln B}
\>,
$$
with the exact two-loop distribution obtained with the EVENT2 
program \cite{CS}.

To obtain the $\as^2$ contribution from our calculation one needs 
to expand eqs.~(\ref{Tfact},\ref{Wfact}) up to second order in $\as$
and multiply by \eqref{Cval}. At order $\as^2$ the result should
accurately reproduce the coefficient of terms  $\as^2\ln^m1/B$ with
$m=2,3,4$. To verify this, in fig.~\ref{fig:event2}, we plot as a
function of $B$ the  differences
$$
\delta_i^{\mathrm{s-log}}(B) \equiv 
   \frac {d\>I^{2\ell}_i(B)}{d\ln B}
  -\frac {d\>I^{s-log}_i(B)}{d\ln B}
\,,
$$
$$
\delta_i^{\mathrm{d-log}}(B) \equiv 
   \frac {d\>I^{2\ell}_i(B)}{d\ln B}
  -\frac {d\>I^{d-log}_i(B)}{d\ln B}
\,,
$$
with $I^{2\ell}_i(B)$ obtained from EVENT2,
$I^{\mathrm{d-log}}_i(B)$ obtained from the expansion of the resummed
result to double-logarithmic accuracy ($\as^2\ln^m1/B$ with $m=3,4$)
and $I^{\mathrm{s-log}}_i(B)$ to single logarithmic accuracy
($\as^2\ln^m1/B$ with $m=2,3,4$). The comparison is made within the
$\MSbar$ scheme, and with $I$ normalised to the Born cross section
rather than the total cross section (as this is what is supplied by
the EVENT2 program). We show only the coefficient of the
$(\as\cf/2\pi)^2$ part of $\delta_i$, as this is the only component
modified by our new treatment of quark recoil.

\FIGURE{
    \epsfig{file=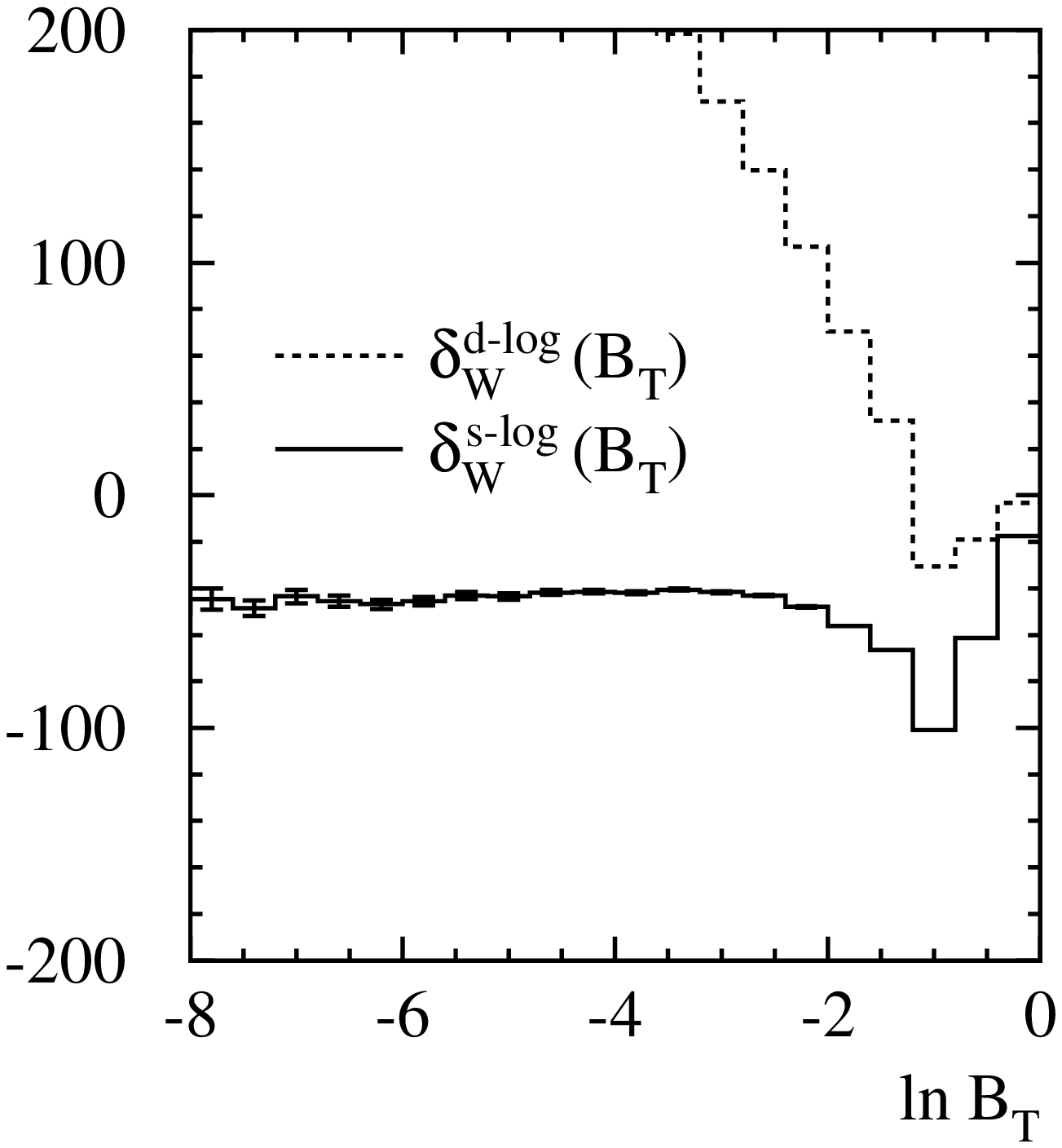,width=0.48\textwidth}
    \epsfig{file=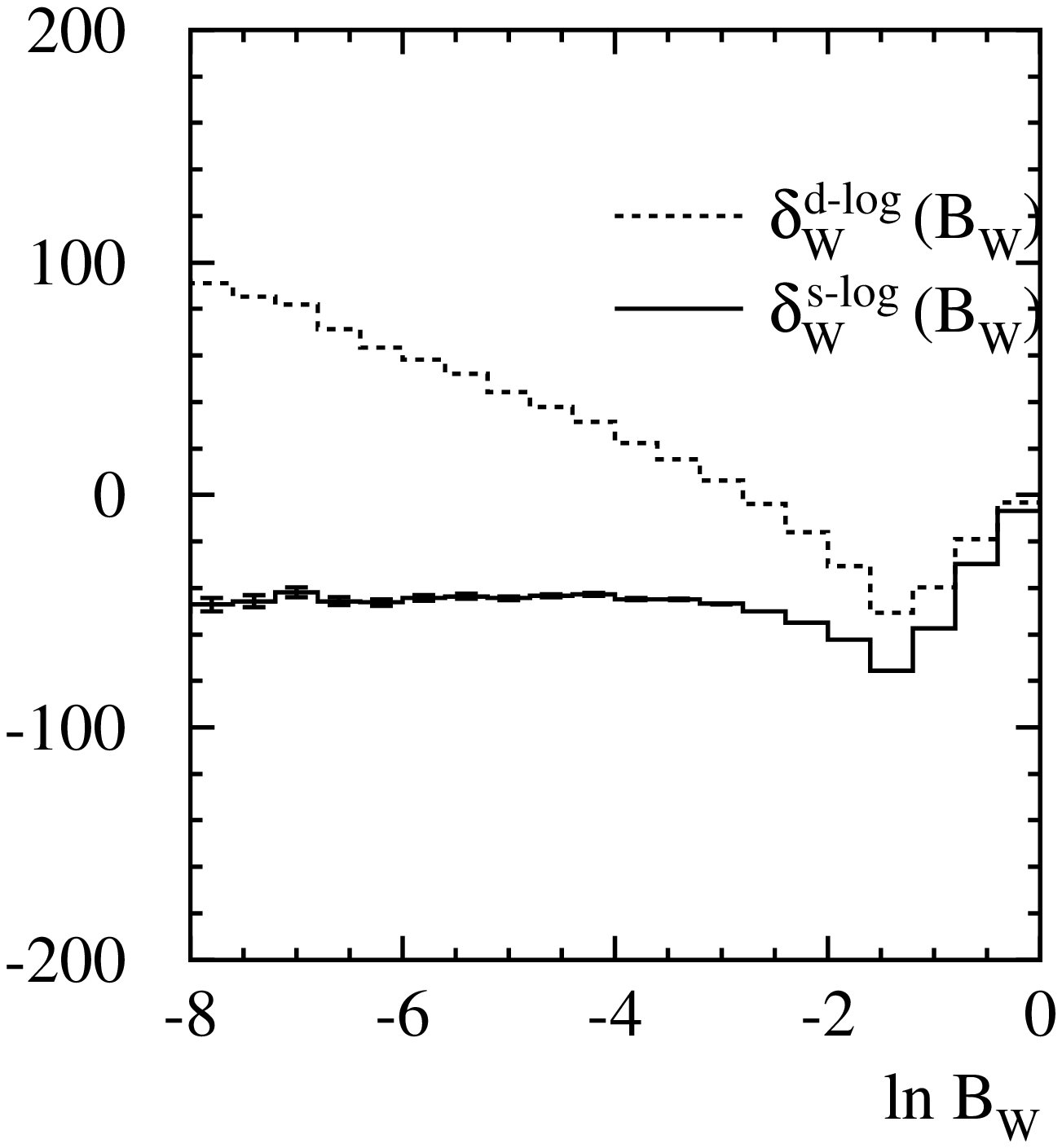,width=0.48\textwidth}
    \caption{The coefficient of the $(\as\cf/2\pi)^2$ component of the
      difference, $\delta_i(B_i)$, between the numerical two-loop
      calculation of the $B$-distribution, performed with EVENT2, and
      the expansion to single and double logarithmic accuracies of the
      resummed result.  Shown for the total and wide broadenings.}
    \label{fig:event2}}

The quantity $\delta_i^{d-log}(B)$ is of order
$\as^2\ln1/B$, corresponding to the absence in
$I^{\mathrm{d-log}}_i(B)$ of the $\as^2\ln^2 1/B$ term. Meanwhile,
$\delta_i^{s-log}(B)$ is of order $\as^2$, indicating that in $I$ we
have correctly taken into account the $\as^2\ln^2 1/B$ term.

\subsection{Matching}
When comparing with experimental data one often chooses to match the
resummed calculation with a full second order calculation (as for
example from the EVENT2 program). Here we give only a brief summary of
two main matching procedures (log-$R$ and $R$ matching schemes) and
refer the reader to \cite{CTTW2} for detailed information. 

For convenience we write
\begin{equation}
\begin{align}
  \label{Gnm}
  \ln \Sigma &= \sum_{n=1}^{\infty} \sum_{m=1}^{n+1} G_{nm}
  \as^n L^m \\
  &= Lg_1(\as L) + g_2(\as L) + \as g_3(\as L) + \cdots,
\end{align}
\end{equation}
with $L=\ln 1/B$. The resummation procedure provides $g_1$, $g_2$ and
$C_1$. Given the resummed calculation, the full two-loop calculation
contains information on $G_{21}$ (the first term of $g_3$) and $C_2$
and on a ``remainder'' which does not necessarily exponentiate.
Matching schemes put together these different parts with the principle
ambiguity being the treatment of the remainder: the log-$R$ scheme
makes the approximation that it exponentiates, the $R$-scheme that it
doesn't.

Here we provide the information that is needed for implementing the
matching with our new calculation. We recommend the use of the second
of the forms in (\ref{Tfact},\ref{Wfact}), using as $\Sigma^{(0)}$ the
equations (18--22) of \cite{CTW}. The factor by which it is
multiplied is
\begin{equation}
  \label{FactorMatching}
  \left(\frac{2}{\lambda}\right)^{2R'} = 
  \left[\int_1^\infty \frac{dx}{x^2} \left(\frac{1+x}{4}\right)^{-R'}
  \right]^2,
\end{equation}
and one should use 
\begin{equation}
  \label{RpMatching}
  R' = \frac{2\as \cf}{\pi} \frac{\ln(1/B)}{1 - 2\alpha_s \beta_0
    \ln(1/B)},
\end{equation}
with $\beta_0 = (11\ca - 2\nf)/12\pi$. This particular form ensures
that the resummed calculation contains only $g_1$ and $g_2$ and no
spurious $g_3$ type terms, as is required for correct matching.

In addition the matching schemes require $G_{1m}$, $G_{2m}$, $C_1$ and
$C_2$ explicitly. $G_{1m}$, $G_{23}$ and $C_1$ are correctly
reproduced in 
\cite{OPAL}\footnote
{Note that the coefficients in \cite{OPAL} 
are given for
  $(\as/2\pi)^n$; this differs from the convention used here.}. 
The new analytic forms for $G_{22}$ are  
\begin{equation}
  \label{G22T}
  G_{22}= \frac1{(2\pi)^2}\left[
    -\left(32\,\ln^2 2+\frac{8}{3}\pi^2\right)\,\cf^2\, 
    + \left(\frac{2}{3}\pi^2-\frac{35}{9}\right)
    \,{\ca}\,{\cf} +\frac{2}{9}\,{\cf}\,{n_f}
  \right]
\end{equation}
for the total broadening, and
\begin{equation}
  \label{G22W}
  G_{22}= \frac1{(2\pi)^2}\left[
    -(32\,\ln^2 2)\,\cf^2\, 
    + \left(\frac{2}{3}\pi^2-\frac{35}{9}\right)
    \,{\ca}\,{\cf} +\frac{2}{9}\,{\cf}\,{n_f}
  \right]
\end{equation}
for the wide broadening.

Finally for the $R$-matching procedure one requires explicit values
for $G_{21}$ and $C_2$. These can be determined from the calculation
performed with EVENT2.  First one determines $G_{21}$ by fitting the
$\as^2\ln(B)$ component of $I(B)$; then one subtracts it out and fits
the asymptotically constant part of $I(B)$, which yields $C_2$. One
obtains:
\begin{center}
  \begin{tabular}{|c|c|c|} \hline 
    & $B_T$       & $B_W$ \\ \hline 
    $G_{21}$ & $1.988\pm0.25$ & $1.869 \pm 0.25$ \\ \hline 
    $C_2$    & $2.330\pm0.25$ & $2.946 \pm 0.25$ \\ \hline
  \end{tabular}
\end{center}
It is necessary to go to extremely small $B\sim e^{-8}$ before a
sufficiently asymptotic behaviour sets in. As a result the value of
$C_2$ which one obtains is strongly dependent on the value taken for
$G_{21}$. The results shown for $C_2$ were obtained by fixing $G_{21}$
(at its central value), and the error is that purely from the fitting
procedure for $C_2$. Roughly, the value of $C_2$ that would have been
obtained with a different $G_{21}$ would be shifted by
$(-8\pm2)[G_{21}-G_{21}(\mathrm{central})]$.

A final aspect of matching that needs to be mentioned is that one
usually introduces a parameter $B_\mathrm{lim}$ at which the resummed
calculation is constrained to vanish through a substitution of the
form 
\begin{equation}
  \frac1{B} \to \frac1{B} - \frac1{B_{\lim}} + 1\,.
\end{equation}
For the purposes of fitting experimental results, a reasonable choice of
$B_{\lim}$ is important; in the section that follows though, it well
be set equal to 1 in order to facilitate the comparison with
results that do not include matching. 

\subsection{The $B_T$ distribution}
To illustrate the changes introduced by our new calculation, we show
in figure~\ref{fig:btshape}a the $B_T$ distribution in the
approximation that $\lambda=2$ and in the case with the full treatment
of the quark recoil. Examining first the case without matching, one
sees that the distribution is shifted towards lower values of $B$, as
expected. Associated
with the shift is an increase 
in the height of the peak of the distribution at small $B$.

If one considers instead the results with log-$R$ matching, one finds
that the effect of going from $\lambda=2$ to full recoil is much
smaller\footnote
{In contrast it is not consistent to show results
  for $\lambda=2$ with $R$-matching, because the $R$-matching
  procedure requires the input of values for $G_{21}$ and $C_{2}$ ---
  these cannot be reliably obtained if one uses a wrong value of $G_{22}$.
  The log-$R$ matching procedure
  doesn't suffer from this problem because it doesn't require the
  explicit input of $G_{21}$ and $C_2$.
  }. 
The explanation is linked to the following property of
log-$R$ matching procedure: if one supplies it with the wrong forms for
$G_{22}$ and $g_2$, as long as those forms are consistent with each
other, the input from the full 2-loop result reestablishes the correct
value for $G_{22}$. 

\FIGURE{
    \epsfig{file=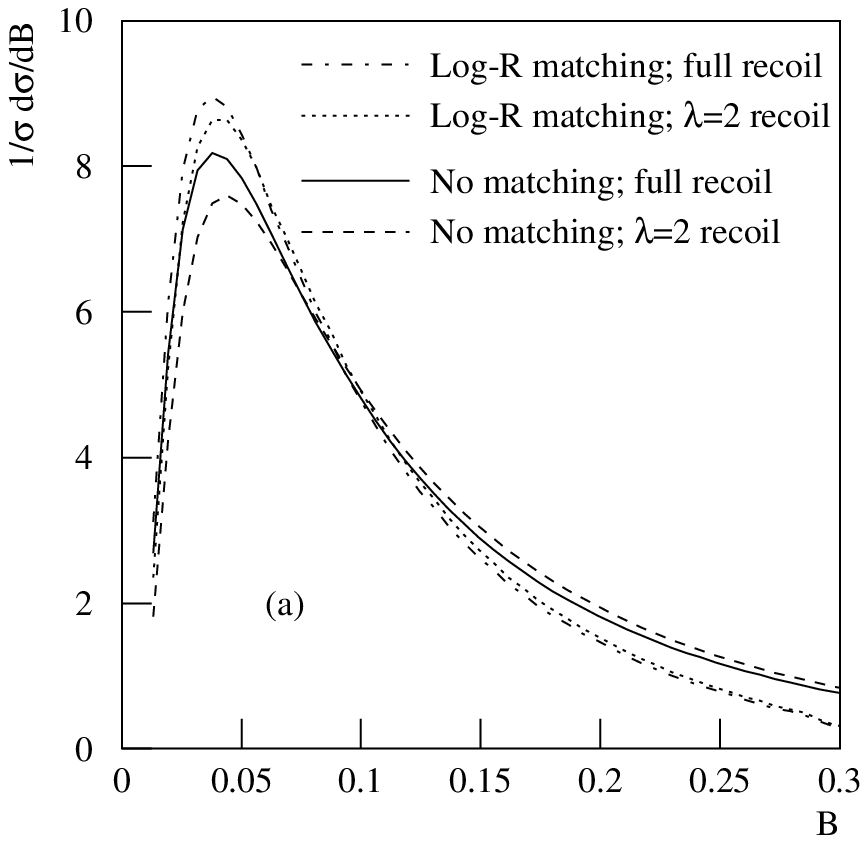,width=0.48\textwidth}
    \epsfig{file=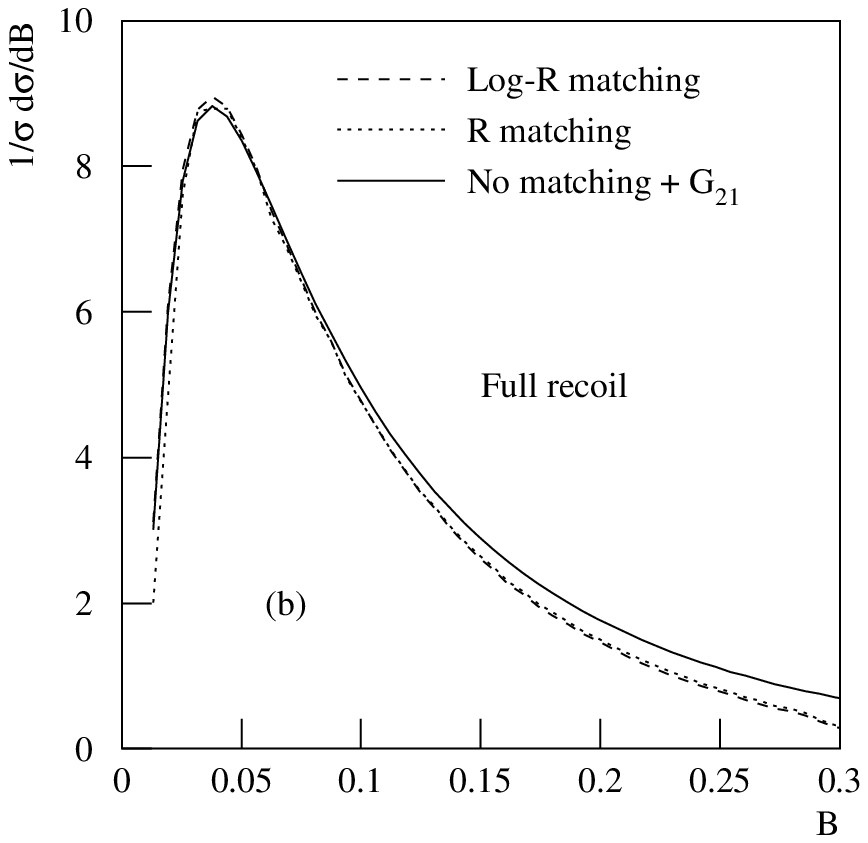,width=0.48\textwidth}
    \caption{(a) The effect of going from the $\lambda=2$
      approximation for the quark recoil, to the full treatment of
      quark recoil, both in the case without any matching and in the
      case of log-$R$ matching; (b) The effect of different matching
      procedures and a comparison with the case without any matching,
      but with the inclusion of the $G_{21}$ term; for all curves,
      $\sqrt{s} = 91.2$~GeV and $\as(M_Z) = 0.118$.}
    \label{fig:btshape}}

In figure~\ref{fig:btshape}a one notes that the matched and
non-matched curves are quite significantly different. At large $B$
this is to be expected, since one is beyond the range of validity of
the resummed calculation. At small $B$ the difference is less welcome.
However there remain certain logarithmic exponentiated terms which
have not been taken into account in our resummed calculation, in
particular those that contribute to $G_{21}$. In
figure~\ref{fig:btshape}b we plot our non-matched resummed calculation
with the additional inclusion of the effect of $G_{21}$ as determined
numerically in the previous section. At small $B$ this leads to very
good agreement with the log-$R$ matched results. At large $B$, as one
expects, a significant difference remains.
Figure~\ref{fig:btshape}b also shows a curve resulting from
$R$-matching, and it is seen to be everywhere in good agreement with
the log-$R$ matching.

\section{Conclusions}

We have shown that a proper treatment of the contribution to $B$ of
the primary quarks recoiling against an ensemble of soft gluons is essential 
for predicting the $B$-distribution with single-logarithmic accuracy. 

We have verified that the evolution equation for the $B$-distribution
derived by CTW embodies all the necessary ingredients to provide the resummed
perturbative prediction with single-logarithmic accuracy. 
The improvement we made concerns the solution of this equation. 
We suggested two forms of the final result, 
\eqref{CTWshift}  and \eqref{CTWfact}. 
The former is the CTW-spectrum evaluated at a rescaled value of 
$B\to 2B/\lambda$, with $\lambda$ a single-logarithmic function, 
$\lambda=\lambda(\as\ln1/B)$ which decreases with $B$ from 2 to 1. 
The latter form is the CTW-answer supplied with the single-logarithmic
factor which does the same job of shifting the distribution to smaller
$B$ values. 
This tendency is opposite to that expected from the $1/Q$ power
correction effect in the $B$-spectrum~\cite{DLMSuniv}.

We were able to show the absence of 
the non-logarithmic contribution proportional to the coupling at
the reduced momentum scale, $\as(BQ)$, which, if present, could damage   
the perturbative prediction at small values of $B$.

It should be noticed, however, that beyond the first order in
$\as(BQ)$, given the present state of the art, such damage looks
unavoidable.  Indeed, consider the most interesting feature of the
$B$-distribution which is its characteristic maximum at
$B=B_{max}\ll1$.  A maximum emerges as a result of an interplay
between the first-order peaked cross section, $\propto(\as\ln1/B)/B$,
and the all-order Sudakov suppression exponent,
$\exp(-R(\lambda/2B))\sim\exp(-\as\ln^21/B)$.  The latter factor takes
over, clearly, when $R'(1/B)\sim \alpha_s(BQ)\ln 1/B$ approaches
unity.  With $B$ decreasing $\ln 1/B$ increases and so does the
running coupling $\as(BQ)$. Formally speaking, in perturbation theory,
that is for $Q\to \infty$, the expansion parameter $\as(BQ)$ stays
small in the region of the maximum.  However, in reality (and for any
foreseeable energies) $\as(B_{max}Q)$ becomes numerically large.  This
undermines the reliability of the perturbative prediction for
$B<B_{max}$ since the neglected subleading corrections of the order
$\as^2\ln1/B$, and among those, $\as(BQ)R'\sim\as(BQ)$ are no longer
numerically negligible.  Corrections of this sort arise, in
particular, from the hard-emission subtraction term $R'\sim
\as\ln1/B\to\as(\ln1/B-3/4)$ in the single-logarithmic pre-exponent.

To trigger genuine confinement effects, the $(A_1\ln Q+A_2)/Q$ shift
in the $B$-distribution \cite{???,DLMSuniv}, it is tempting to look
at smaller $Q$ values.  However, some care should be exercised here
since the kinematical range of $B$ shrinks.  Moreover, one should bear
in mind the above-mentioned intrinsic uncertainty of the perturbative
prediction
for $B<B_{max}$ which becomes larger for smaller values of $Q$.  To be
on a safe side, one should try to stay with the $B$-values to the
right of the maximum, $B>B_{max}$.

It remains to be seen whether the $1/Q$ power correction extracted
with the use
of the improved perturbative expression derived in the present paper,
will exhibit the expected $\ln Q$ enhancement.

\acknowledgments
We benefited much from continuous discussions of
  this and related subjects with Bryan R. Webber, Michael H. Seymour
  and Stefano Catani. 

\appendix

\section{One-gluon emission with single-logarithmic accuracy.}

The single gluon emission distribution is
$$
\frac{d\sigma_1}{\sigma}=
dx_1dx_2 \frac{\as \cf}{2\pi}\frac{x_1^2+x_2^2}{(1-x_1)(1-x_2)}\,
\qquad x_1=\frac{2pQ}{Q^2}\,, \quad x_2=\frac{2\bar pQ}{Q^2}\,.
$$
We introduce the gluon c.m. energy fraction $z$ and transverse
momentum $k_t$
$$
z=2-x_1+x_2\,, 
\quad\quad 
\frac{k_t^2}{Q^2}=\frac{(1-x_1)(1-x_1)}{1-z}\equiv \epsilon^2\,.
$$
If the gluon is in the right hemisphere we have
$$
x_1=1-\half\left(z+\sqrt{z^2-4(1-z)\epsilon^2}\right)\,,
\quad
x_2=1-\half\left(z-\sqrt{z^2-4(1-z)\epsilon^2}\right)\,.
$$
If the gluon is in the left hemisphere 
these expressions are interchanged.

In terms of the gluon variables we have 
$$
\frac{d\sigma_1}{\sigma}=
  2\,\frac{d^2k_{t}}{\pi k_{t}^2} dz\>
  \frac{\as \cf}{2\pi}\>
  \frac{1 + (1-z)^2 -2(1-z)\epsilon^2}{\sqrt{z^2 -4(1-z)\epsilon^2}}\,,
$$
where the factor $2$ takes into account that the gluon can be emitted
in the right or left hemispheres. 
Integrating,
$$
\int_{z_0}^1 dz  
\frac{1 + (1-z)^2 -2(1-z)\epsilon^2}{\sqrt{z^2 -4(1-z)\epsilon^2}}
=2\ln \frac{1}{\epsilon}-\frac32 
-4\epsilon^2\ln \epsilon + \cO{\epsilon^4}\,,
$$
where $z_0$ is the zero of the square root and is given by 
$z_0=2\epsilon +\cO{\epsilon^2}$.
Then for small $\epsilon$ we can approximate the quark splitting
function by replacing 
$$
\frac{1 + (1-z)^2 -2(1-z)\epsilon^2}{\sqrt{z^2 -4(1-z)\epsilon^2}}
\>\Rightarrow\>  
\frac{1 + (1-z)^2}{z}\>\Theta(z - \frac{k_t}{Q})\,,
$$
where the leading correction is of order $\epsilon^2\ln \epsilon$.
It produces a contribution $\cO{\as(Q)}$ to the integral of the gluon 
emission and is taken care of by the factor $C(\as(Q))$.

\section{The integral {\protect \eqref{sigma'}}.}

The coefficient $\Delta(\nu,b)$ of $R''(\mu)$ in the expansion of 
$\delta R(\nu,b)$ in \eqref{sigma'} is given by
\begin{equation}
\begin{split}
  \Delta(\nu,b) =
  -\frac{d}{d\eps}
  \left(\frac{1-(\mu/\nu)^\eps\,\Gamma(1+\eps)}{\eps}\right) +
  \ln\mu\ln\frac{\nu}{\mu} +c(\nu,b)\\
  = \half\left(\ln\frac\mu\nu+\gamma_E\right)^2
  + \frac{\pi^2}{12}\>+\> \ln\mu\ln\frac{\nu}{\mu}\>+\>c(\nu,b) \,,
\end{split}
\end{equation}
with $c(\nu,b)$ given by
\begin{equation}
  \label{cdef}
\begin{split}
  c(\nu,b)
= \int_0^{\infty}\frac{dx}{x}\>\ln x\>
  e^{-\nu x}\left(J_0(bx)-1\right)\>
= \ln \nu \ln \frac{\mu}{\nu}
+\int_0^{\infty}\frac{dx}{x}\>\ln x\>
  e^{-x}\left(J_0(xb/\nu)-1\right)\>.
\end{split}
\end{equation}
This shows that $\Delta(\nu,b)$ depends only on the ratio $\mu/\nu$. 

For the relevant integrals over $b$ we use
\begin{align}
  \int_0^\infty \frac{\nu bdb}{(\nu^2+b^2)^{3/2}}
  &= \int_1^\infty \frac{dy}{y^2}= 1\>, \\
  \int_0^\infty \frac{\nu bdb}{(\nu^2+b^2)^{3/2}} \cdot \ln\frac{\mu}{\nu}
  &= \int_1^\infty \frac{dy}{y^2}\ln\frac{1+y}{2}\> =\> \ln 2\>, \\
  \int_0^\infty \frac{\nu bdb}{(\nu^2+b^2)^{3/2}} \cdot
  \ln^2\frac{\mu}{\nu}
  &= \int_1^\infty \frac{dy}{y^2}\ln^2\frac{1+y}{2}\> =\>
  \frac{\pi^2}{6} - \ln^22\>.
\end{align}
The inverse Fourier integral of $c(\nu,b)$ is
\begin{multline}
 \int_0^\infty \frac{\nu bdb}{(\nu^2+b^2)^{3/2}}\>c(\nu,b)
 = \int bdb\int pdp e^{-\nu p} J_0(bp)
 \int_0^{\infty}\frac{dx}{x}\>\ln  x\>
  e^{-\nu x}\left(J_0(bx)-1\right) \\
= \int_0^{\infty}\frac{dx}{x}\>\ln  x\>
  \left[\,e^{-2\nu x}-e^{-\nu x}\,\right]
= -\gamma_E \ln2 +\half\ln^22 +\ln2\ln\nu \>.
\end{multline}
Assembling all terms, we obtain the result in the text.





\newpage
\noindent

\begin{thebibliography}{9}
\bibitem{Broad}
   P.E.L. Rakow and B.R. Webber, \npb{191}{1981}{63}; \\
   R.K. Ellis and B.R. Webber, 
   in Proc. 1986 Summer Study on the Physics of the Superconducting 
   Supercolliders, Snowmass, Colorado, 1986, ed. R.~Donaldson and
   J.~Marx (Division of Particles and Fields of the APS, New York, 1987).  
\bibitem{???}
 Yu.L. Dokshitzer and  B.R. Webber,  \plb{404}{1997}{321}.

\bibitem{DLMSuniv}
  Yu.L.  Dokshitzer, A. Lucenti,  G. Marchesini and G.P. Salam,
  IFUM-601-FT; \\
  M. Dasgupta and B.R. Webber,   in preparation.
\bibitem{CTW}
     S. Catani, G. Turnock and B.R. Webber, \plb{295}{1992}{269}.
\bibitem{CS}
   S. Catani and M. Seymour, \npb{485}{1997}{291}.
\bibitem{CMW}
   S. Catani, G. Marchesini and B.R. Webber,  \npb{349}{1991}{635} ;\\
   Yu.L. Dokshitzer, V.A. Khoze and S.I. Troyan, \prd{53}{1996}{89}.

\bibitem{CTTW2}
     S. Catani, L. Trentadue, G. Turnock and B.R. Webber,
     \npb{407}{1993}{3}.
\bibitem{OPAL} The OPAL collaboration, \zpc{59}{1993}{1}.
\end{thebibliography}
\end{document}